\documentclass[aps,pre,superscriptaddress,amsmath,amssymb,twocolumn]{revtex4}
\usepackage{graphicx}
\usepackage{amsmath}
\usepackage{hyperref}
\usepackage{color}
\definecolor{darkblue}{RGB}{8,81,156}
\hypersetup{colorlinks,breaklinks,
            linkcolor=darkblue,urlcolor=darkblue,
           anchorcolor=darkblue,citecolor=darkblue}

\usepackage{array}
\newcolumntype{L}[1]{>{\raggedright\let\newline\\\arraybackslash\hspace{0pt}}m{#1}}
\newcolumntype{C}[1]{>{\centering\let\newline\\\arraybackslash\hspace{0pt}}m{#1}}
\newcolumntype{R}[1]{>{\raggedleft\let\newline\\\arraybackslash\hspace{0pt}}m{#1}}

\date{\today}

\allowdisplaybreaks

\begin{document}

\title{Computational investigation of structure, dynamics and nucleation kinetics of a family of modified Stillinger-Weber model fluids in bulk and free-standing thin films}

\author{Melisa M. Gianetti}
%\email{mgianetti@qi.fcen.uba.ar}
\affiliation{
DQIAQF/INQUIMAE-CONICET, Facultad de Ciencias Exactas y Naturales, Universidad de Buenos Aires,  Ciudad Universitaria, Buenos Aires, Argentina
%Departamento de Qu\'imica Inorg\'anica, Anal\'itica y Qu\'imica F\'isica (DQIAQF) / Instituto de Qu\'imica F\'isica de los Materiales, Medio Ambiente y Energ\'ia (INQUIMAE-CONICET), Facultad de Ciencias Exactas y Naturales, Universidad de Buenos Aires, Pabell\'on II, Ciudad Universitaria, (1428), Buenos Aires, Argentina
}

\author{Amir Haji-Akbari}
%\email{hajakbar@princeton.edu}
\affiliation{Department of Chemical and Biological Engineering, Princeton University, Princeton NJ 08544}

\author{M. Paula Longinotti}
%\email{longinot@qi.fcen.uba.ar}
\affiliation{
DQIAQF/INQUIMAE-CONICET, Facultad de Ciencias Exactas y Naturales, Universidad de Buenos Aires,  Ciudad Universitaria, Buenos Aires, Argentina
%Departamento de Qu\'imica Inorg\'anica, Anal\'itica y Qu\'imica F\'isica (DQIAQF) / Instituto de Qu\'imica F\'isica de los Materiales, Medio Ambiente y Energ\'ia (INQUIMAE-CONICET), Facultad de Ciencias Exactas y Naturales, Universidad de Buenos Aires, Pabell\'on II, Ciudad Universitaria, (1428), Buenos Aires, Argentina
}

\author{Pablo G. Debenedetti}
\email{pdebene@exchange.princeton.edu}
\affiliation{Department of Chemical and Biological Engineering, Princeton University, Princeton NJ 08544}

\date{\today}

\begin{abstract}
In recent decades, computer simulations have found increasingly widespread use as powerful tools of studying phase transitions in wide variety of systems. In the particular and very important case of aqueous systems, the commonly used force-fields tend to offer quite different predictions with respect to a wide range of thermodynamic and kinetic properties, including the ease of ice nucleation, the propensity to freeze at a vapor-liquid interface, and the existence of a  liquid-liquid phase transition. It is thus of fundamental and practical interest to understand how different features of a given water model affect its thermodynamic and kinetic properties.  In this work, we use the forward-flux sampling technique to study the crystallization kinetics of a family of modified Stillinger-Weber (SW) potentials with energy ($\epsilon$) and length ($\sigma$) scales taken from the monoatomic water (mW) model, but with different tetrahedrality parameters ($\lambda$). By increasing $\lambda$ from 21 to 24, we observe the nucleation rate increases by 48 orders of magnitude at a  supercooling of ${\zeta}=T/T_m=0.845$.  Using classical nucleation theory, we are able to demonstrate that this change can largely be accounted for by the increase in $|\Delta\mu|$, the thermodynamic driving force. We also perform rate calculations in freestanding thin films of the supercooled liquid, and observe a crossover from a surface-enhanced crystallization at $\lambda = 21$ to a bulk-dominated crystallization for $\lambda\ge22$. 
\end{abstract}

\maketitle

\section{Introduction\label{section:introduction}}

Water is one of the most ubiquitous substances on earth. In spite of its abundance and importance, however, many questions about it warrant further scrutiny. One of the most notable examples is  ice nucleation, which, despite its relevance to areas such as biology and meteorology, is far from fully understood. Ice nucleation plays a very important role in the atmosphere, and its kinetics affect the modulation of solar radiation and hydrological fluxes in the atmosphere~\cite{ZenderJGeoPhys1994, HeggRepProgPhys2009, MurrayGRL2015}. Due to the exponential dependence of nucleation rate on temperature, it is only possible to make rate measurements across narrow ranges of temperature~\cite{ButorinKristallografiya1972, AndersonJAtmosSci1980, HagenJAtmosScie1981, TaborekPRB1985, SassenJAtmosSci1988, KramerJCP1999, LeisnerJPCA2005, NillsonJPCL2015}, with extrapolations to other temperatures prone to large uncertainties~\cite{HajiAkbariPNAS2015}. Furthermore, the microscopic time and length scales relevant to ice nucleation are not accessible to experiments. This makes obtaining mechanistic information about freezing a very challenging task with existing experimental techniques. 

%It is, therefore, not possible to obtain any mechanistic information about freezing from such nucleation experiments, at least for the time being.

In the absence of such high-resolution experiments, computer simulations are attractive alternatives as, by construction, they provide a detailed microscopic perspective of the nucleation process. However, computational studies of nucleation are very challenging and for realistic molecular models of water, the direct observation of homogeneous nucleation of ice in the absence of biasing potentials and external stimuli was achieved only recently~\cite{Matsumoto2002}. Since then, freezing molecular dynamics (MD) trajectories have been obtained for a variety of force fields~\cite{JungwithJPCB2006, JungwirthJMolLiq2007, JungwirthJPhysChemC2010, MolineroPCCP2011, YagasakiPRE2014}. Nevertheless, the statistical nature of freezing could not be properly sampled in those studies as only a few freezing trajectories were obtained. On the other hand, there is a large body of work involving the use of biasing potentials along pre-chosen reaction coordinates to drive crystallization and  map its free energy landscape.~\cite{TroutJACS2003, BrukhnoJPhysCondMat2008, QuigleyJCP2008, ReinhardtJCP2012, PalmerNature2014}. Despite their utility in estimating quantities such as nucleation barriers, these bias-based techniques are not suitable for studying the kinetics of nucleation as they distort the underlying dynamics of the  system. An unbiased statistically relevant approach would require collecting a large number of trajectories, an undertaking that is almost impossible with regular molecular dynamics simulations of realistic molecular models of water. 

A more practical alternative is to use state-of-the-art  path sampling techniques that sample the reactive trajectories in a targeted manner. One such technique is forward flux sampling (FFS)~\cite{AllenFrenkel2006}, a powerful algorithm that has been recently used for studying a wide range of first-order transitions such as hydrophobic evaporation~\cite{SumitPNAS2012}, droplet coalescence~\cite{FrenkelJCP2007}, wetting~\cite{SavoyLangmuir2012}, magnetic switching~\cite{VoglerPRB2013}, protein folding~\cite{EscobedoJCP2007} and crystallization~\cite{SanzPRL2007, LiNatMater2009, GalliJChemPhys2009, GalliPCCP2011, GalliNatComm2013, ThaparPRL2014, HajiAkbariFilmMolinero2014, HajiAkbariPNAS2015}. Li \emph{et al.}~and Haji-Akbari \emph{et al.} employed this technique to study the kinetics of ice nucleation for the mW coarse-grained model of water, both in the bulk~\cite{GalliPCCP2011, HajiAkbariFilmMolinero2014} and in confined geometries~\cite{GalliNatComm2013, HajiAkbariFilmMolinero2014}. Recently, Haji-Akbari and Debenedetti used a coarse-grained variant of FFS to perform the first direct calculation of nucleation rate for a molecular model of water~\cite{HajiAkbariPNAS2015}, in this particular case the TIP4P/Ice model~\cite{VegaTIP4PiceJCP2005}. Galli and collaborators have also used FFS to study nucleation kinetics in other tetrahedral liquids such as Si and Ge~\cite{LiNatMater2009, GalliJChemPhys2009}. Since FFS yields a large number of trajectories, the statistical nature of crystallization can be studied and precise nucleation rates can been obtained. In addition, such computational studies can  be used for deducing useful mechanistic information about freezing~\cite{HajiAkbariPNAS2015}. They can also be utilized as a comparative tool, in order to determine the effect of a design parameter, or an external stimulus on the kinetics and mechanism of nucleation~\cite{LiNatMater2009, GalliJChemPhys2009, GalliNatComm2013, HajiAkbariFilmMolinero2014}.

A notable example of applying FFS as a comparative tool is the quest for determining the role of a vapor-liquid interface on freezing, a question regarded as one of the ten most important unanswered questions about ice~\cite{RauschNature2013}. In a seminal paper~\cite{TabazadehPNAS2002}, Tabazadeh \emph{et al.} proposed that a vapor-liquid interface will enhance the crystallization of liquids such as water that partially wet their crystal. This has steered an ongoing controversy involving both  experimental~\cite{DuftACPD2004, ShawJPCB2005, SignorellPhysRevE2008, GurganusJPhysChemLett2011} and  computational~\cite{JungwithJPCB2006, JungwirthJMolLiq2007, JungwirthJPhysChemC2010, GalliNatComm2013, HajiAkbariFilmMolinero2014} studies. Jungwirth \emph{et al.}~\cite{JungwithJPCB2006, JungwirthJMolLiq2007, JungwirthJPhysChemC2010} 
used conventional molecular dynamics simulations to study free-standing thin films of a six-site model of water~\cite{NadaJCP2003} and observed an enhancement of crystallization close to the vapor-liquid interface. They attributed this enhancement to the lack of electrostatic neutrality close to the interface, leading to the emergence of a net electric field in the subsurface region. Electric fields are known to stimulate crystallization in water~\cite{KusalikPRL1994, KusalikJACS1996}. Subsequently, Li~\emph{et al.} used FFS rate calculations to demonstrate that crystallization is favored at  free interfaces of the tetrahedral liquids Si and Ge~\cite{LiNatMater2009, GalliJChemPhys2009}. They argued that density fluctuations at the interface facilitate the crystallization of liquids that are denser than their corresponding crystals. Accordingly, they hypothesized that surface-induced crystallization would also be observed for water, which also possesses a negatively-slope melting curve. In the case of water nano-droplets simulated using the coarse-grained mW potential~\cite{GalliNatComm2013}, however, they observed a significant reduction in the nucleation rate with respect to the bulk, an observation rationalized by the fact that those nano-droplets were subject to large Laplace pressures. Haji-Akbari \emph{et al.}~\cite{HajiAkbariFilmMolinero2014} used  conventional molecular dynamics simulations, FFS and umbrella sampling to study ice nucleation in freestanding thin films of mW water that, by construction, have zero Laplace pressure. However, they observed that crystallization was suppressed at the vapor-liquid interface. They attributed their observations to the fact that the nuclei that emerge in the interfacial region are more aspherical than their bulk counterparts. This work was an unequivocal counter-example to both the negative-slope melting curve  and the partial wettability hypotheses, as the mW model satisfies both these criteria.

These apparent inconsistencies are partly due to the sensitivity of nucleation kinetics to thermodynamic features of the underlying force fields, such as the presence or absence of electrostatic interactions.  Due to recent advances in computer architecture, and in molecular simulation techniques, it is now possible to systematically estimate the characteristic relaxation and nucleation times for different water models. In this context, there is a large variability. Some models such as TIP4P/Ice~\cite{VegaTIP4PiceJCP2005} and TIP4P/2005~\cite{VegaTIP4P2005} almost never crystallize in molecular dynamics simulations, and there is an astronomical separation of relaxation and nucleation time scales in conventional MD simulations~\cite{HajiAkbariPNAS2015}.  At the other end of the spectrum, are the models such as the monoatomic water (mW) model~\cite{MolineroJPCB2009} that spontaneously crystallize at sufficiently low temperatures. This  coarse-grained model of water was developed by re-parameterizing the Stillinger-Weber potential~\cite{StillingerPRB1985} that had been originally developed for group IV elements. 

Another difference  between different water models, is the presence of a second liquid-liquid critical point. The mW model does not exhibit a liquid-liquid transition in the supercooled regime~\cite{LimmerJCP2011}, while this transition is observed in the molecular ST2 model~\cite{PalmerNature2014}. At present, it is not fully understood what features of these different water models lead to such stark differences in the separation of structural relaxation and crystallization time scales, or in the existence or absence of a second critical point. A systematic approach for addressing these fundamental questions is to investigate the sensitivity of these specific predictions to particular features of the underlying water-like models. In this work, we are interested in this very fundamental question, and we investigate the role of the tetrahedrality of a family of  coarse-grained water models on the nucleation kinetics by studying the SW potentials with different tetrahedrality parameters. We are, in particular, interested in the effect of the tetrahedrality parameter on:~(i) the nucleation kinetics, and (ii) the suppression or facilitation of crystallization at vapor-liquid interfaces.

This paper is organized as follows. Section~\ref{section:methods},  Methods, is divided into four subsections. In Section~\ref{section:sw}, we introduce the family of SW potentials considered in this work. Section~\ref{section:md} and \ref{section:ffs} are dedicated to technical details of the molecular dynamics simulations and the FFS calculations, respectively. The particular order parameter used for tracking the progress of crystallization is discussed in Section~\ref{section:op}. Section~\ref{section:results},  Results and Discussion, is  divided into three subsections. In Section~\ref{section:summary}, the  rate calculations are summarized. In Section~\ref{section:bulk}, we provide an in-depth analysis of the dependence of bulk nucleation rates on $\lambda$, the tetrahedrality parameter, while in Section~\ref{section:interface}, the effect of $\lambda$ on the facilitation or suppression of crystallization at a vapor-liquid interface is discussed. Finally, Section~\ref{section:conclusions} is reserved for our concluding remarks.

\section{Methods\label{section:methods}}
\subsection{The Stillinger-Weber potential\label{section:sw}}

In this work, we consider a family of tetrahedral liquids, described by the Stillinger-Weber potential~\cite{StillingerPRB1985}:
\begin{eqnarray}
E &= & \sum_i\sum_{j>i} \phi_2(r_{ij}) + \sum_i\sum_{j\neq i}\sum_{k<j}\phi_3(\textbf{r}_{ij},\textbf{r}_{ik},\theta_{ijk})
\end{eqnarray}
with
\begin{eqnarray}
\phi_2(r) & =& A\epsilon\left[B\left(\frac{\sigma}{r}\right)^p-\left(\frac{\sigma}{r}\right)^q\right]\psi(r)\label{eq:two-body}\\
\phi_3(r,s,\theta) &=& \lambda\epsilon[\cos\theta-\cos\theta_0]^2\psi(r)\psi(s)\label{eq:three-body}\\
\psi(r)&=& \exp\left[\frac{\sigma}{r-a\sigma}\right]
\end{eqnarray}
Here $r_{ij}$ is the distance between the particles $i$ and $j$ and $\theta_{ijk}$ is the angle between the $\textbf{r}_{ij}$ and $\textbf{r}_{ik}$ displacement vectors. $\phi_2(r)$ corresponds to the two-body interactions between two individual particles, while $\phi_3(r,s,\theta)$ is a three-body term that is used for enforcing tetrahedrality in the liquid. The values of $A$, $B$, $p$, $q$, $\gamma$ and $a$ are constant for different parameterizations of the SW potential, and are given in Table~\ref{tab:sw:parameters}. The exponential terms in Eqs.~(\ref{eq:two-body}) and (\ref{eq:three-body}) are to assure that both the potential and its derivatives go to zero at $r=a\sigma$.
 The tetrahedrality parameter, $\lambda$, modulates the energetic penalty of deviating from $\theta_0= 109.47^{\circ}$, the ideal tetrahedral angle. In the original parameterization of the SW potential for Group IV elements, $\lambda$ is 20, 21 and 26.2 for Ge, Si, and C, respectively. In the SW-based water potential, mW, $\lambda=23.15$. In this work, we investigated tetrahedral liquids with $\lambda = 21, 22, 23.15$ and 24, with $\epsilon = 6.189$~kcal/mol and $\sigma = 2.3925$~\AA~taken from the mW potential~\cite{MolineroJPCB2009}.

\begin{table}
	\centering
	\caption{\label{tab:sw:parameters}Parameters of the Stillinger-Weber potential}
	\begin{tabular}{C{2.0cm}C{2.0cm}C{2.0cm}C{2.0cm}}
	\hline\hline
	Parameter & Value & Parameter & Value\\
	\hline
	$A$ & 7.049556277 & $p$ & 4 \\
	$B$ & 0.6022245584  & $q$ & 0 \\
	$\gamma$ & 1.2 & $a$ & 1.8 \\
	\hline
	\end{tabular}
\end{table}

\subsection{System Preparation and Molecular Dynamics Simulations\label{section:md}}

Molecular dynamics simulations are performed using LAMMPS~\cite{PimptonLAMMPS1995}. Newton's equations of motion are integrated using the velocity-Verlet algorithm~\cite{SwopeJCP1982} with a time step of 2~fs. Temperature is controlled using a Nos\'{e}-Hoover~\cite{NoseMolPhys1984, HooverPhysRevA1985} thermostat with $\tau = 0.2$~ps. In $NpT$ simulations, pressure is controlled using a Parinello-Rahman barostat~\cite{ParrinelloJAppPhys1981} with $\tau = 2.0$~ps.

Throughout this study, we carry out two types of MD simulations, all in boxes that are periodic in all three directions. For the bulk geometry, simulations are carried out in the isothermal-isobaric, $NpT$, ensemble. Initial configurations are obtained by quenching and compressing a dilute simple cubic lattice of $4,\!096$ molecules to the target temperature and pressure. These simulations are carried out for a minimum of 40~ns, which is much longer than the characteristic structural relaxation times of all the systems considered in this work. The characteristic relaxation times, as computed from the decay of the self-intermediate scattering function, are in the order of a few picoseconds for all the systems considered in this work~\cite{HajiAkbariPNAS2015, HajiAkbariFilmMolinero2014}. For the  film geometry, simulations are carried out in the isothermal-isochoric, $NVT$, ensemble. In this case, the cubic boxes of the configurations equilibrated in the bulk geometry are expanded along the $z$ direction by a factor of four, and the resulting configurations are equilibrated in $NVT$ MD simulations for an additional 40 ns. The resulting films are roughly 5-nm in thickness, and the expansion of the box along the $z$ direction assures that the films are not affected by their periodic images.  For rate calculations, all simulations are carried out at a relative supercooling of, ${\zeta} = T/T_m= 0.845$ with $T_m$, the equilibrium melting temperatures at zero pressure obtained from Ref.~\cite{MolineroJPCB2009} and given in Table~\ref{table:Tm}. The rationale behind fixing ${\zeta}$, and not $\Delta T=T_m-T$, will be explained in Section~\ref{section:bulk}.

\begin{table}[h]
	\centering
	\caption{\label{table:Tm}Equilibrium melting temperatures for different $\lambda$'s, obtained from Ref.~\cite{MolineroJPCB2009}.}
	\begin{tabular}{C{2cm}C{3cm}}
	\hline\hline
	$\lambda$ & $T_m$~[K] \\
	\hline
	21 & 206\\
	22 & 240\\
	23.15 & 274\\
	24 & 291\\
	\hline\hline
	\end{tabular}
\end{table}

\subsection{Forward-flux Sampling\label{section:ffs}}
Nucleation rates are computed using the forward-flux sampling technique~\cite{AllenFrenkel2006}. In this method, the process of transitioning from $A$, the metastable liquid basin, to $B$, the crystalline basin is simulated in stages defined by an order parameter, $\xi$,  that evolves monotonically between the two basins. Individual stages are defined with the milestones $\xi_A<\xi_0<\xi_1<\cdots<\xi_B$ with each stage involving the sampling of trajectories that start at $\xi=\xi_i$, and cross the next milestone, $\xi=\xi_{i+1}$, or return to $\xi=\xi_A$, the original basin. Before starting the first stage, regular MD simulations are carried out in the $A$ basin in order to gather  sufficient number of configurations at $\xi_0$, the first milestone after the basin. The positioning of the individual milestone is so that the transition probability between successive milestones is between $10^{-3}$ and $10^{-1}$, except for large $\xi$'s (close to $\xi_B$) for which the transition probabilities are close to unity. The FFS calculation is terminated when $P(\xi_i|\xi_{i-1})=1$ for every $\xi_i>\xi_{i-1}$. The nucleation rate is then given by:
\begin{eqnarray}
R &=& \Phi_0\prod_{i=1}^NP(\xi_{i}|\xi_{i-1})
\end{eqnarray}
Here, $\Phi_0$ is the flux of trajectories that cross $\xi_0$ after leaving $A$ (see below), and $\prod_{i=1}^NP(\xi_{i}|\xi_{i-1})$   is the probability that a trajectory initiated from a configuration at $\xi_0$ reaches the crystalline basin, $(\xi\ge\xi_B)$. The flux, $\Phi_0$, is computed from a series of MD simulations in $A$ using the following expression:
\begin{eqnarray}
\Phi_0 &=& \frac{N_0}{t\langle V\rangle}
\end{eqnarray}
Here, $N_0$ is the total number of crossings (of $\xi_0$) originating in $A$, $\langle V\rangle$ is the average volume of the liquid and $t$ is the length of the underlying MD trajectory. For the bulk geometry, $\langle V\rangle$ is the average volume of the simulation box. For thin films, $\langle V\rangle$ is obtained by partitioning the simulation box into a grid of cubic cells of size 3.2~\AA~and  enumerating the average number of cells that have at least eleven nonempty neighboring cells~\cite{HajiAkbariFilmMolinero2014}. For both geometries, these simulations are continued until a minimum of 700 configurations are obtained at $\xi_0$. 

To compute $P(\xi_{i+1}|\xi_i)$, the transition probability from $\xi_i$ to $\xi_{i+1}$, a configuration at $\xi_i$ is randomly chosen, and its momenta are randomized using the Maxwell-Boltzmann distribution. The arising MD trajectory is  integrated until it 'succeeds` by crossing $\xi_{i+1}$, or it fails  by returning back to $\xi_A$. This procedure is repeated until a minimum of 700 configurations are collected at $\xi_{i+1}$. $P(\xi_{i+1}|\xi_i)$ is then estimated as the fraction of successful trajectories.

\subsection{Order Parameter\label{section:op}}
\begin{figure}[h]
\centering
\includegraphics[width=.45\textwidth]{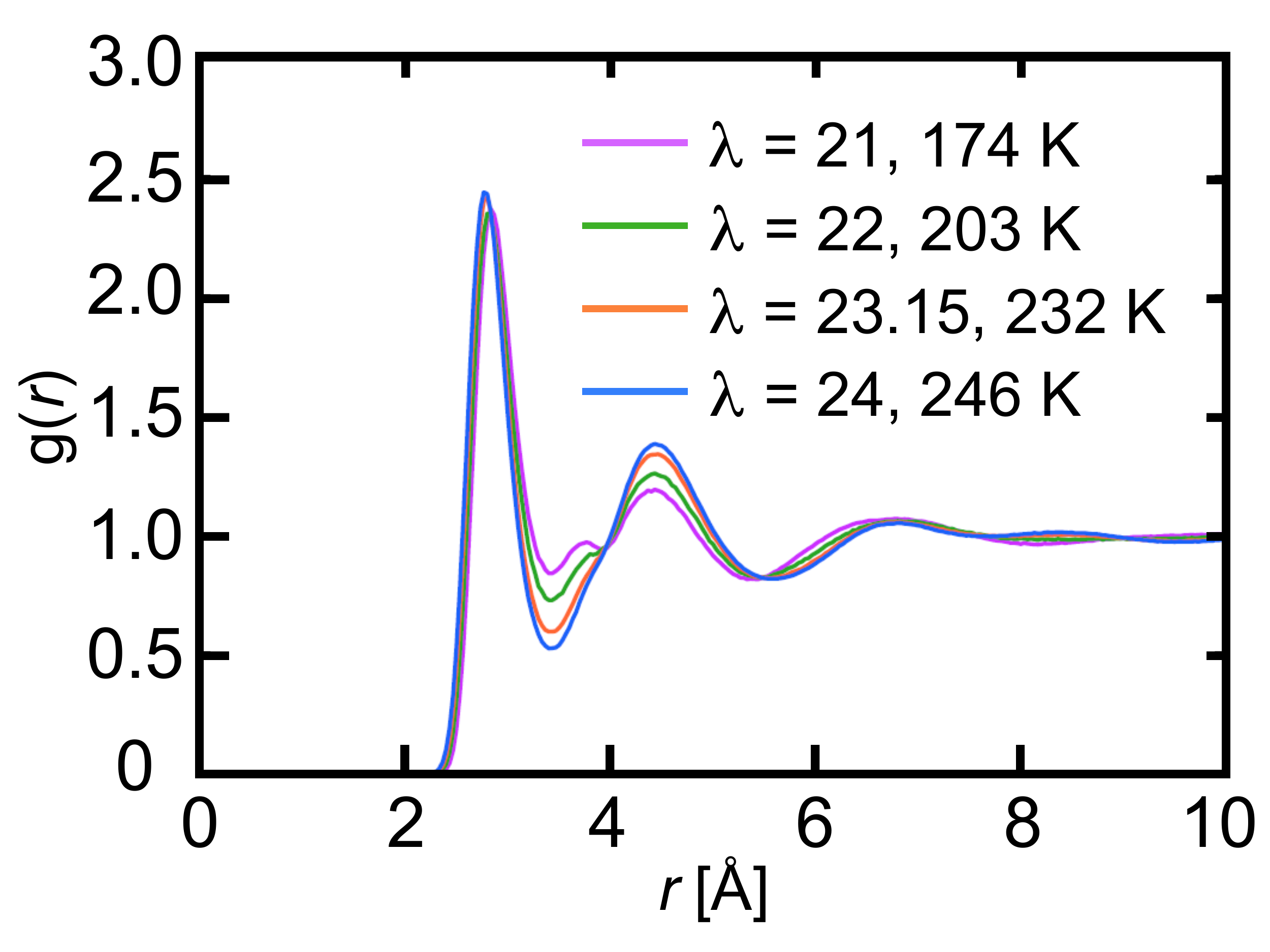}
\caption{\label{fig:RDF}Radial distribution functions computed for the modified SW potentials at $p= 1$~bar and ${\zeta}=0.845$.}
\end{figure}

As has been customary in crystallization studies~\cite{LiNatMater2009, GalliJChemPhys2009, GalliPCCP2011, GalliNatComm2013, HajiAkbariFilmMolinero2014, HajiAkbariPNAS2015}, $\xi$ is chosen to be the number of molecules in the largest solid-like cluster in the system. To this end, all solid- and liquid-like molecules in the system are detected using the $q_6$ bond-orientational order parameter, proposed by Steinhardt~\emph{et al.}~\cite{SteinhardtPRB1983}. For each molecule, $i$, $q_{6m}(i)$ is given by:
\begin{eqnarray}
q_{6m}(i) &=& \frac{1}{N_b(i)}\sum_{j=1}^{N_b(i)} Y_{6m}(\theta_{ij}, \phi_{ij})
\end{eqnarray}
with $N_b(i)$ the number of nearest neighbors of $i$ (as defined per a distance cutoff) and $\theta_{ij}$ and $\phi_{ij}$ denote the azimuthal and polar angles associated with the displacement vector $\textbf{r}_{ij}$. A scalar invariant of $q_{6m}(i)$ is given by:
\begin{eqnarray}
q_6(i) &=& \frac{1}{N_b(i)}\sum_{j=1}^{N_b(i)}\frac{\textbf{q}_6(i)\cdot\textbf{q}_6^*(j)}{|\textbf{q}_6(i)||\textbf{q}_6(j)|}
\end{eqnarray}
In accordance with earlier studies~\cite{GalliPCCP2011, GalliNatComm2013, HajiAkbariFilmMolinero2014, HajiAkbariPNAS2015}, we use a distance cutoff of $r_c = 3.2$~\AA~for $\lambda = 24$. For $\lambda = 21$ and 22, a larger nearest neighbor shell is utilized, with a distance cutoff of 3.45~\AA. As shown in Fig.~\ref{fig:RDF}, the first neighbor shell expands as $\lambda$ decreases.  For $\lambda = 21$ and 22, FFS calculations never converge when a distance cutoff of 3.2~\AA~is utilized.

\begin{figure}
\centering
\includegraphics[width=.45\textwidth]{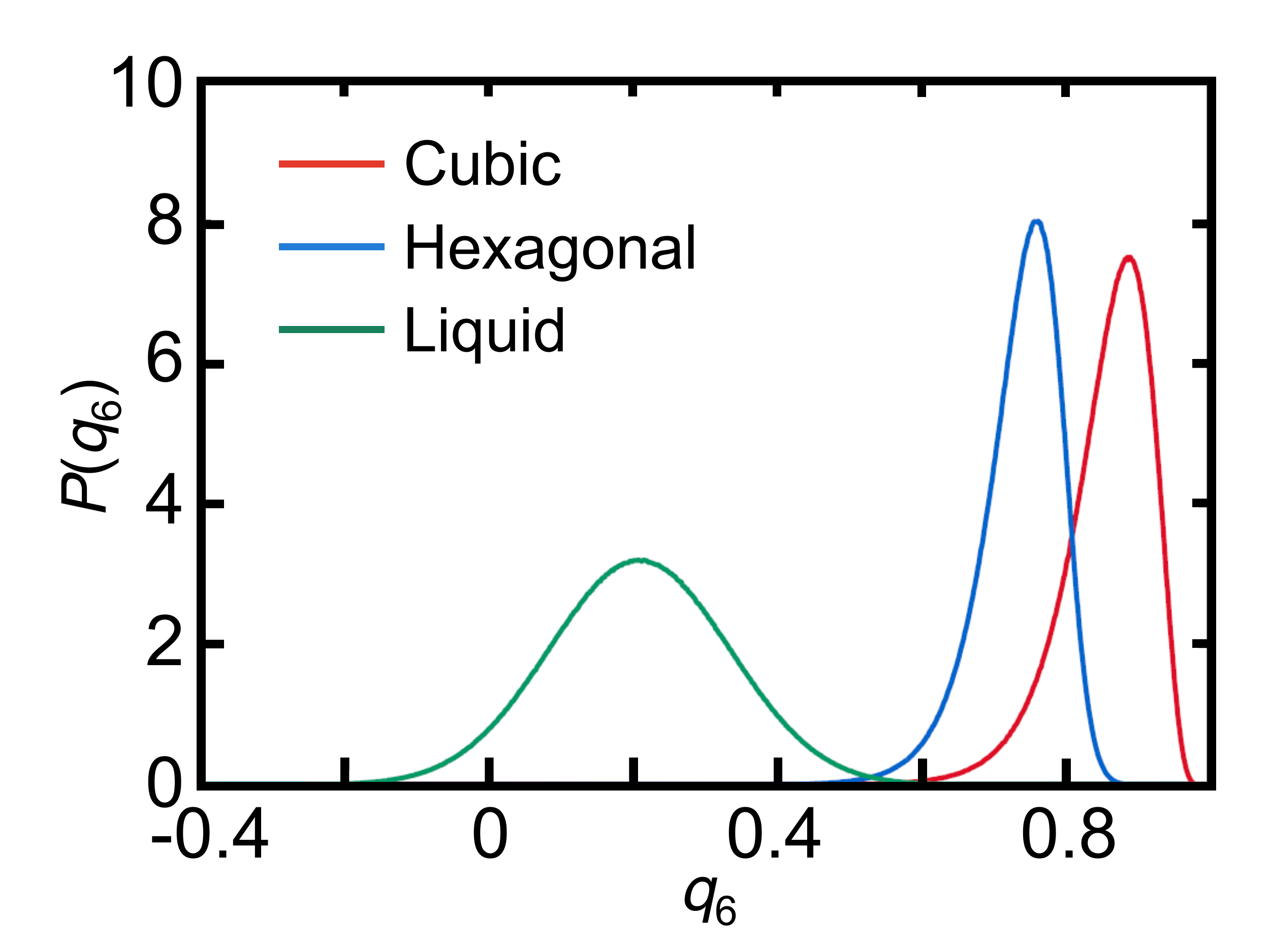}
\caption{\label{fig:q6} $q_6$ distributions for the cubic and hexagonal crystals and the supercooled liquid for $\lambda=22$ at $T=203$~K and $p=1$~bar. }
\end{figure}

Fig.~\ref{fig:q6} depicts the $q_6$ histograms for the supercooled liquid, and the cubic and hexagonal crystals at $\lambda=22$ and $T=203$~K. Similar histograms are observed for the other $\lambda$ values investigated in this work. Note that there is very little overlap between the $q_6$ distribution in the liquid and the crystal, making $q_6$ a robust way of distinguishing solid- and liquid-like molecules. Accordingly, every molecule with $q_6>0.5$ is labelled as solid-like, and the solid-like molecules that are within the nearest neighbor shell of one another are grouped together to form clusters of solid-like molecules.  In order to remove chains of locally tetrahedral molecules and to obtain more compact crystallites, we apply the chain exclusion algorithm developed by Reinhardt \emph{et al.}~\cite{VegaJCP2012} to the resulting clusters.

\section{Results and Discussions\label{section:results}}
\subsection{Summary of Nucleation Rate Calculations\label{section:summary}}
\begin{figure}[h]
\centering
\includegraphics[width=.42\textwidth]{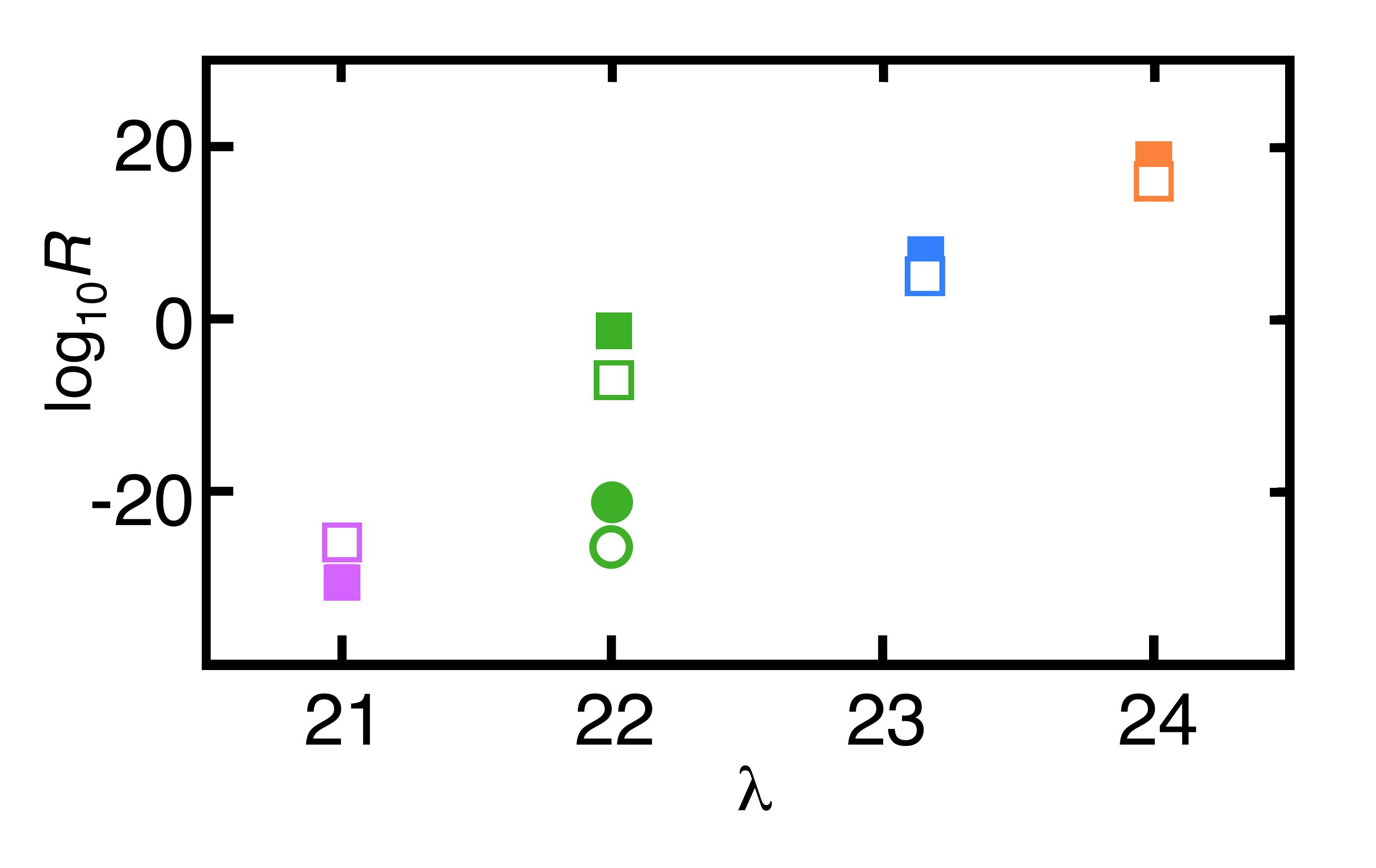}
\caption{\label{fig:rates}Nucleation rates for different $\lambda$'s and geometries. All bulk calculations were performed at 1~bar. Values for 23.15 are taken from Ref.~\cite{HajiAkbariFilmMolinero2014}. Filled and empty symbols correspond to the bulk and film geometries, respectively. Squares and circles correspond to ${\zeta}=0.845$ and ${\zeta}=0.87$, respectively.}
\end{figure}

Table~\ref{table:rates} and Fig.~\ref{fig:rates} summarize the rates computed from FFS. For $\lambda=23.15$, rates are obtained from Ref.~\cite{HajiAkbariFilmMolinero2014}. For $\lambda=22$, we carry out an additional set of rate calculations at $T=209$~K for reasons that will be explained in Section~\ref{section:interface}. This corresponds to a relative supercooling of ${\zeta}=0.87$, slightly higher than the ${\zeta}$ for other systems. 

Table~\ref{table:basin} summarizes the technical specifications of the basin simulations.  Note that $\Phi_0$ is fairly insensitive to $\lambda$. This is not surprising considering that $\Phi_0$ strongly depends on the particular choice of the $\xi_A$ and $\xi_0$ milestones. Indeed, we always choose $\xi_0$ towards of the 0.1\% tail of the $\xi$ histogram in the liquid basin.  Considering the thermal nature of fluctuations in $\xi$, such a uniform criterion will give rise to fluxes that are more or less of the same order of magnitude.

\begin{table*}
\centering
\caption{\label{table:rates}Computed fluxes, cumulative transition probabilities, and nucleation rates for different $\lambda$'s and geometries. Error bars are computed using the procedure described in Ref.~\cite{AllenFrenkel2006}.}
\begin{tabular}{C{1cm}C{1.7cm}C{2.0cm}C{3.6cm}R{3.6cm}R{3.6cm}}
\hline\hline
$\lambda$ & $T[\text{K}]$ & Geometry & $\log_{10}\Phi_0 [\text{m}^{-3}\cdot\text{s}^{-1}]$ & $\log_{10}P(\xi_B|\xi_0)$ & $\log_{10}R [\text{m}^{-3}\cdot\text{s}^{-1}]$\\
\hline
21 & 174 & Bulk & $+34.578\pm0.016$ & $-64.921\pm0.565$ & $-30.343\pm$0.565 \\
21 & 174 & Film & $+35.416\pm0.005$ & $-61.652\pm0.576$ & $-26.236\pm0.576$\\
22 & 203 & Bulk & $+35.357\pm0.014$ & $-36.782\pm0.414$ & $-1.423\pm0.414$ \\
22 & 203 & Film & $+35.584\pm0.011$ & $-42.475\pm0.504$ & $-6.891\pm0.504$\\
22 & 209 & Bulk & $+34.378\pm0.021$ & $-55.682\pm0.497$ & $-21.304\pm0.497$\\
22 & 209 & Film & $+35.285\pm0.008$ & $-61.714\pm0.605$ & $-26.429\pm0.605$\\
24 & 209 & Bulk & $+36.020\pm0.008$ & $-17.468\pm0.294$ & $+18.552\pm0.294$\\
24 & 209 & Film & $+36.057\pm0.007$ & $-19.857\pm0.387$ & $+16.200\pm0.387$\\
\hline
\end{tabular}
\end{table*}

\begin{table*}
\centering
\caption{\label{table:basin}Technical specifications of basin simulations}
\begin{tabular}{C{1cm}C{1.5cm}C{2.5cm}C{1cm}C{1cm}C{1.5cm}C{1.5cm}C{2cm}C{3cm}}
\hline\hline
$\lambda$ & $T$~[K] & Geometry & $\xi_A$ & $\xi_0$ & $N_0$ & $t$~[ns] & $\langle V\rangle$~[nm$^3$] & $\Phi_0~[\text{m}^{-3}\cdot\text{s}^{-1}]$ \\
\hline
21 & 174 & bulk & 1 & 5 & $2,\!887$ & 664.90 & 114.653 & $3.7871\times10^{34}$ \\
21 & 174 & film & 2 & 8 & $27,\!538$ & 755.00 & 139.926 & $2.6066\times10^{35}$\\
22 & 203 & bulk & 2 & 6 & $3,\!883$ & 144.00 & 118.356 & $2.2783\times10^{35}$ \\
22 & 203 & film & 3 & 8 & $5,\!969$ & 127.09 & 122.236 & $3.8422\times10^{35}$\\
22 & 209 & bulk & 2 & 7 & $1,\!654$ & 585.50 & 118.297 & $2.3880\times10^{34}$\\
22 & 209 & film & 3 & 8 & $12,\!092$ & 436.58 & 143.651 & $1.9281\times10^{35}$\\
24 & 246 & bulk & 5 & 10 & $12,\!595$ & 96.00 & 125.830 & $1.048\times10^{36}$\\
24 & 246 & film & 5 & 10 & $16,\!233$ & 111.76 & 127.320 & $1.141\times10^{36}$\\
\hline
\hline
\end{tabular}
\end{table*}

\subsection{$\lambda$ Dependence of Bulk Nucleation Kinetics\label{section:bulk}}

As can be observed in Table~\ref{table:rates} and Fig.~\ref{fig:rates}, the nucleation rate is a strong function of $\lambda$. For a relative supercooling of ${\zeta}=0.845$, $R_{\text{bulk}}$ increases by $\approx48$ orders of magnitude, from $\log_{10}R=-30.343$ at $\lambda=21$ to $\log_{10}R=+18.552$ at $\lambda = 24$. This trend is consistent with our intuition that nucleating a tetrahedral crystal must be easier from a more tetrahedral liquid (i.e.,~a liquid with higher $\lambda$). 

In order to obtain a more quantitative understanding,  we employ the classical nucleation theory (CNT)~\cite{WeberZPhyskCHem1926, BeckerAnnPhys1935, TurnbullJCP1949, PabloBook1996}, which is a particularly useful quantitative framework for studying nucleation.  Despite its approximate nature, CNT provides a physically reasonable picture of the nucleation process and can thus be used not only for making sense of the observed/computed nucleation rates, but also for predicting rates under conditions at which direct rate measurements/calculations are not feasible.  By assuming that a crystalline nucleus is in quasi-equilibrium with the surrounding liquid, CNT yields the following expression for $R$, the nucleation rate:
\begin{eqnarray}
R = A \exp\left[-\frac{\Delta G^*}{k_BT}\right] \label{eq:rate:CNT}
\end{eqnarray}
Here, $A$ is the kinetic pre-factor given by:
\begin{eqnarray}
A &=& \frac{24Z\rho_{\text{liq}}D\xi^{*2/3}}{l^2}\label{eq:pre-factor}
\end{eqnarray}
with $D$ and $\rho_{\text{liq}}$, the self-diffusivity and the density of the liquid. $l$ is the atomic jump distance in the liquid and corresponds to the diffusion mean path. $Z$ is the Zeldovich factor, which depends on both the thermodynamic driving force, $|\Delta\mu|$, and the critical nucleus size $\xi^*$:
\begin{eqnarray}
Z &=& \left(\frac{|\Delta\mu|}{6\pi k_BT\xi^*}\right)^{1/2}\label{eq:zeldovich}
\end{eqnarray}
Here, $|\Delta\mu|$ is the absolute value of the difference between the chemical potentials of the metastable liquid and the stable crystalline phase.
The nucleation rate possesses an exponential dependence on $\Delta G^*$, the nucleation barrier, a quantity that accordingly plays a crucial role in determining the magnitude of the nucleation rate. For a spherical nucleus, $\Delta G^*$ is given by:
\begin{eqnarray}
\Delta G^* &=& \frac{16\pi\gamma_{ls}^3}{3\rho_s^2|\Delta\mu|^2}
\label{eq:barrier}
\end{eqnarray}
Here $\gamma_{ls}$ is the solid-liquid surface tension and $\rho_s$ is the number density of the crystal. Assuming that $\Delta h_m$, the melting enthalpy, and $\Delta s_m$, the melting entropy, are not strong functions of temperature, Eq.~(\ref{eq:barrier}) can be rewritten as:
\begin{eqnarray}
\Delta G^* &=& \frac{16\pi\gamma_{ls}^3}{3\rho_s^2\Delta h_m^2(1-T/T_m)^2}
\label{eq:barrier:simplified}
\end{eqnarray}
Eq.~(\ref{eq:barrier:simplified}) is the main motivation behind choosing ${\zeta}=T/T_m$ over $\Delta T=T_m-T$ as the supercooling parameter in this study. The assumption of  $\Delta h_m$ and $\Delta s_m$ being independent of temperature is, however, not very accurate for water considering its heat capacity anomaly. Therefore, we do not use Eq.~(\ref{eq:barrier:simplified}) for any quantitative analysis.

\begin{figure}
	\centering
	\includegraphics[width=.45\textwidth]{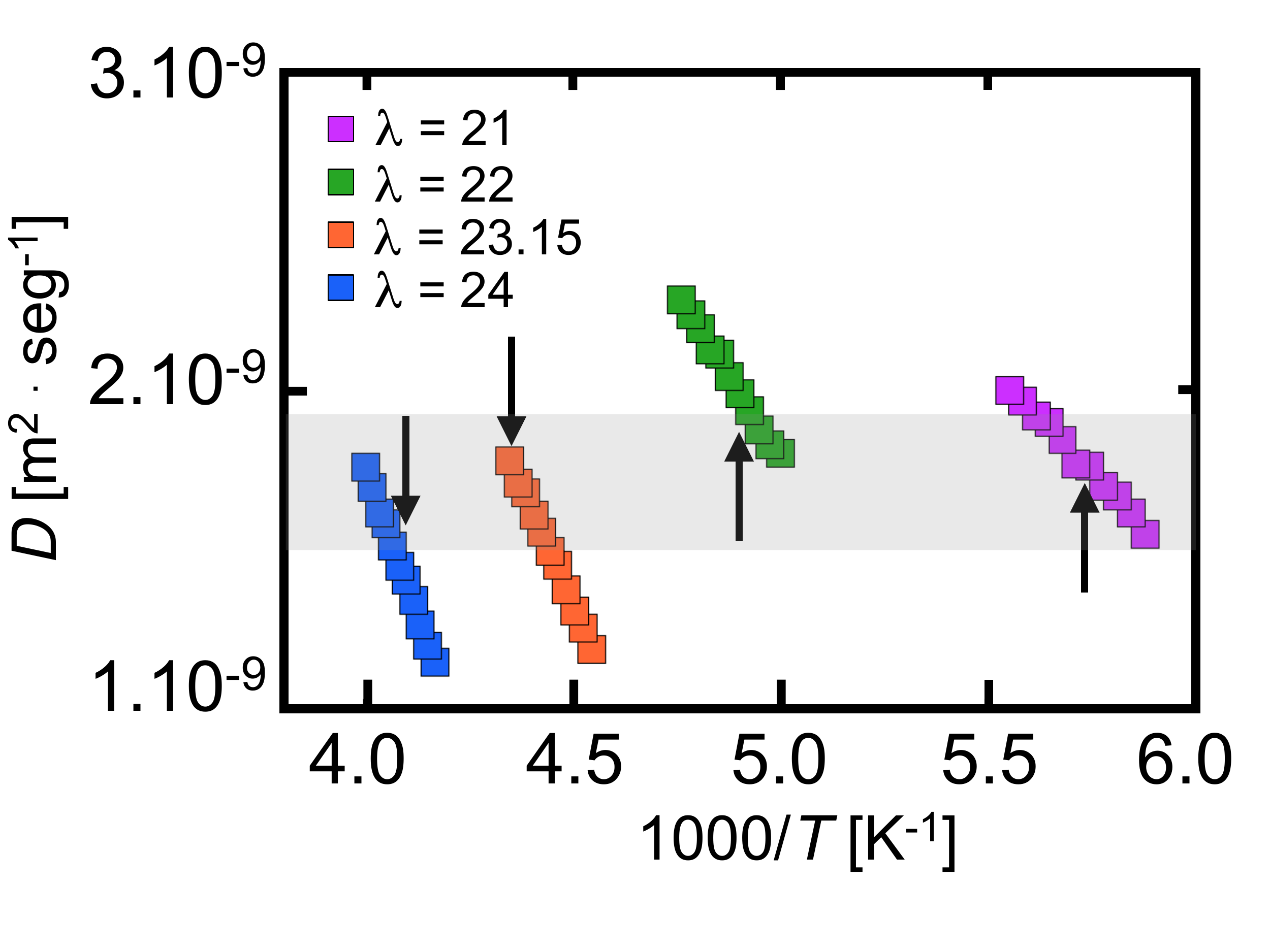}
	\caption{\label{fig:diffusivity} Temperature and $\lambda$ dependence of self-diffusivities in the bulk at $p=1$~bar. Arrows correspond to a supercooling of $\zeta=0.845$. The shaded gray region corresponds to the extent of change in diffusivity upon fixing $\zeta$. }
\end{figure}

Using Eqs.~(\ref{eq:rate:CNT}), (\ref{eq:pre-factor}) and (\ref{eq:barrier}), it is possible to assess the relative contributions of the kinetic pre-factor and the nucleation barrier to the observed changes in $R$. It is clear from Eq.~(\ref{eq:pre-factor}) that the kinetic pre-factor, $A$, is most sensitive to self-diffusivity, $D$. Fig.~\ref{fig:diffusivity} depicts $D$ vs.~$T$ computed from $NpT$ simulations at 1~bar and $0.80<{\zeta}<0.88$ using the well-known Einstein formula~\cite{HelfandPhysRev1960}. At fixed ${\zeta}$, $D$ does not change significantly with $\lambda$, and is always around  $1.7\times10^{-9}\pm 0.3\times10^{-9}~\text{m}^{2}/\text{s}$. This makes $A$, the kinetic pre-factor, insensitive to changes in $\lambda$ at fixed ${\zeta}$. Therefore, any change in $R$ is almost exclusively a consequence of the change in the nucleation barrier. 

It is  interesting to note that a fixed ${\zeta}$ would correspond to a higher absolute temperature in a liquid with higher $\lambda$, and yet, the self-diffusivity remains unchanged. In other words, increasing $\lambda$ is qualitatively equivalent to decreasing temperature. This can be explained by noting that the liquid becomes more structured at higher $\lambda$'s, due to the emergence of more locally tetrahedral arrangements. Such added structuring will make it more difficult for molecules to escape their tetrahedral cages. Therefore, the propensity to form more local tetrahedral arrangements offsets the faster dynamics at higher absolute temperatures, and keeps self-diffusivity almost unchanged as long as $\zeta$ is constant. This picture is consistent with earlier observations that in supercooled water, the low-density liquid, which is highly tetrahedral, has much larger relaxation times- and much smaller diffusivities- in comparison to the less-tetrahedral high-density liquid at identical temperatures~\cite{ErringtonNature2001}. 

\begin{figure}
	\centering
	\includegraphics[width=.45\textwidth]{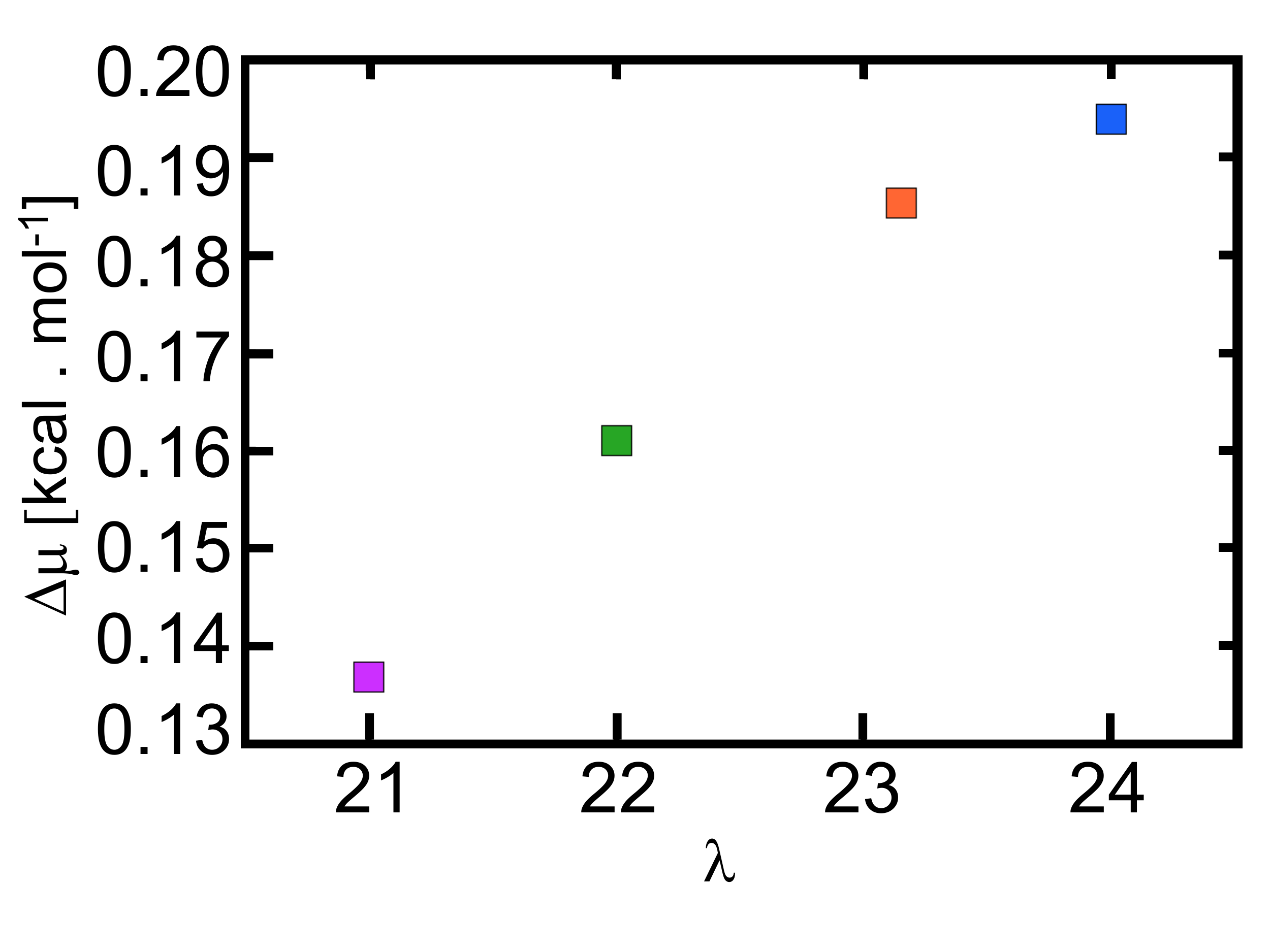}
	\caption{\label{fig:d-mu} $|\Delta\mu|$ as a function of $\lambda$ at ${\zeta}=0.845$. }
\end{figure}

The exponential term in Eq. (\ref{eq:rate:CNT}), however, depends on two thermodynamic properties: $|\Delta\mu|$, the free energy difference between the liquid and the solid, and $\gamma_{ls}$, the liquid-solid surface tension. Slight changes in either of these quantities can alter the nucleation rate by several orders of magnitude. As noted earlier in the literature~\cite{HajiAkbariFilmMolinero2014, HajiAkbariPNAS2015}, it is notoriously difficult to estimate $\gamma_{ls}$ in the supercooled regime, due to the difficulty of stabilizing a solid-liquid interface. On the contrary, $|\Delta\mu|$ is very easy to compute and is obtained from thermodynamic integration:
\begin{eqnarray}
\Delta\mu (T) &=& T\int_{T}^{T_m} \frac{h_{\text{liq}}(\overline{T})-h_{\text{hex}}(\overline{T})}{\overline{T}^2}d\overline{T}
\end{eqnarray}
with $h_{\text{liq}}$ and $h_{\text{hex}}$ the molar enthalpies of the liquid and the hexagonal crystal, respectively. Those enthalpies are obtained from $NpT$ simulations from $\langle h\rangle=\langle u\rangle +p\langle v\rangle$. As depicted in Fig.~\ref{fig:d-mu}, $|\Delta\mu|$ is a strictly increasing function of $\lambda$ at constant ${\zeta}$. This is in line with our intuition that a tetrahedral crystal will become more stable as $\lambda$ increases. However, it is not known \emph{a priori} whether such an increase in $|\Delta\mu|$ is sufficient for explaining the 48 orders of magnitude increase in the nucleation rate.

\begin{figure}
\centering
\includegraphics[width=.45\textwidth]{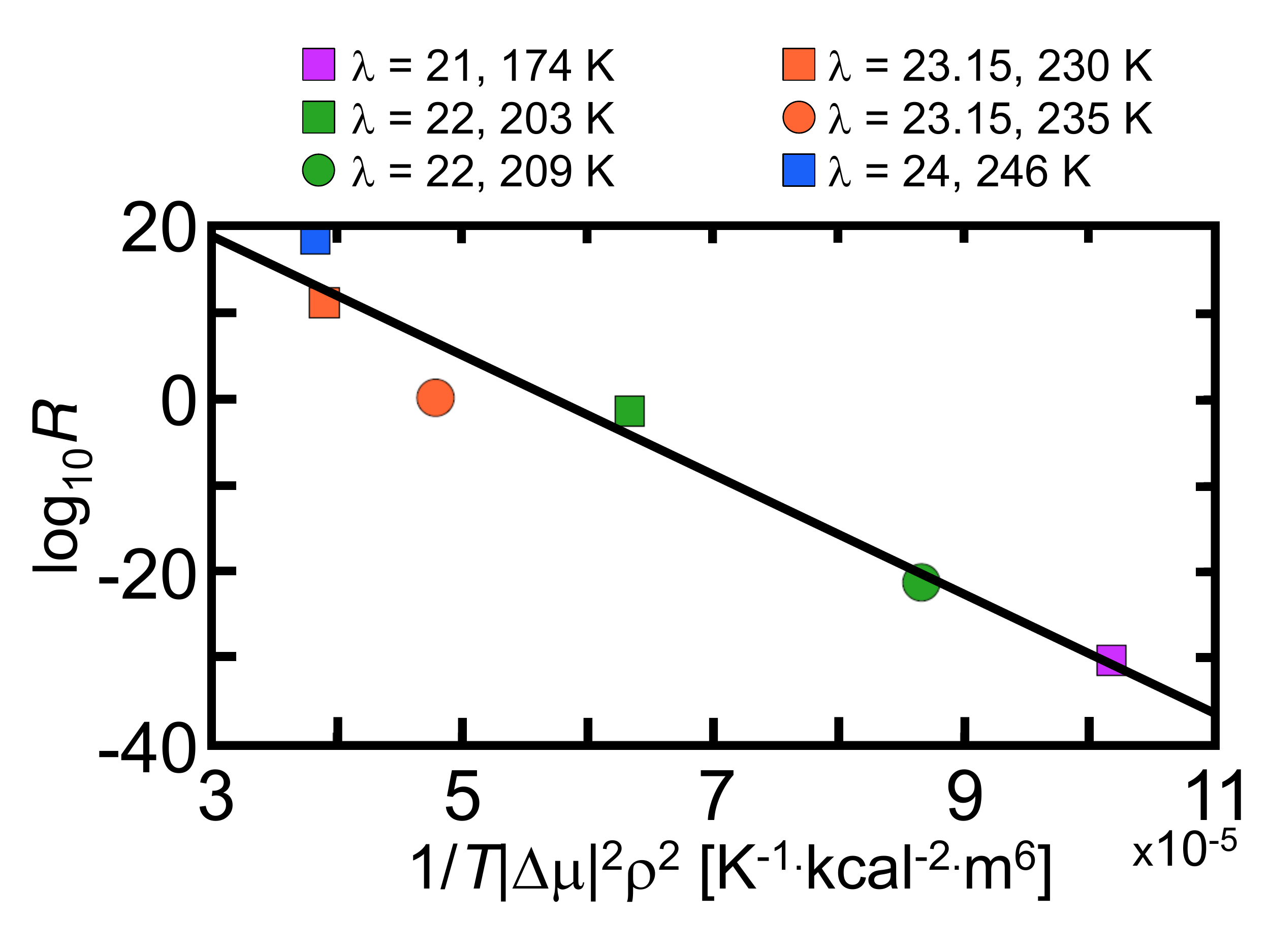}
\caption{\label{fig:CNT:fit} $\log_{10}R$ vs.~$1/(\rho_s^2|\Delta\mu|^2T)$ for the rate calculations in the bulk. The two rates at $\lambda=23.15$ are taken from Ref.~\cite{HajiAkbariFilmMolinero2014}. }
\end{figure}

A systematic way of deciphering the relative contributions of $|\Delta\mu|$ and $\gamma_{ls}$ on $R$ is to assume that $\gamma_{ls}$ is constant and independent of $\lambda$ and temperature. It then follows from Eq.~(\ref{eq:barrier}) that a linear fit must exist between $\log R$ and $1/(\rho_s^2|\Delta\mu|^2T)$:
\begin{eqnarray}
\log_{10}R &=& \log_{10}A-\frac{16\pi\gamma_{ls}^3}{3k_B\ln 10}\cdot\frac{1}{T\rho_s^2|\Delta\mu|^2}\label{eq:linear-fit}
\end{eqnarray}
 Fig.~\ref{fig:CNT:fit} depicts $\log_{10}R$ vs.~$1/(\rho_s^2|\Delta\mu|^2T)$ for the four bulk rate calculations performed in this work as well as the two rate calculations of Ref.~\cite{HajiAkbariFilmMolinero2014} at ${\zeta} = 0.839$ and 0.858. The linear fit is reasonably good with $R^2 = 0.9538$.  It is therefore safe to conclude that the observed change in nucleation rates can, for the most part, be explained by the change in $|\Delta\mu|$. As a corollary, the possible contribution of $\gamma_{ls}$ to the rate is expected to be minimal. Indeed, one expects that $\gamma_{ls}$ will decrease upon increasing $\lambda$, as there will be higher structural similarity between the liquid and the crystal at more tetrahedral liquids.  However, such a change does not appear to be very important, at least over the range of $\lambda$'s considered in this work. It has indeed been observed that $\gamma_{ls}$ is a weak function of temperature for the mW model~\cite{VegaJCP2014}, and our findings suggest that the same assertion might be true for the $\lambda$ dependence of  $\gamma_{ls}$. Indeed, the $\gamma_{ls}$ obtained from Eq.~(\ref{eq:linear-fit}) is $28.14\pm2.95~\text{mN}\cdot\text{m}^{-1}$, which is statistically indistinguishable from earlier estimates of $31.01~\text{mN}\cdot\text{m}^{-1}$~\cite{GalliPCCP2011}, $30~\text{mN}\cdot\text{m}^{-1}$~\cite{LimmerJCP2012} and $29~\text{mN}\cdot\text{m}^{-1}$~\cite{VegaJCP2014} for $\gamma_{ls}$ in the mW system.

\begin{table}
	\centering
	\caption{\label{table:ratios}Ratio between bulk and film nucleation rates for different $\lambda$'s. Values for $\lambda=23.15$ are obtained from interpolating the rates given in Ref.~\cite{HajiAkbariFilmMolinero2014}.}
	\begin{tabular}{C{1cm}C{1cm}C{2.0cm}C{2.5cm}}
	\hline\hline
	$\lambda$ & $T[\text{K}]$ & ${\zeta}=T/T_m$ & $R_{\text{bulk}}/R_{\text{film}}$\\
	\hline
	21 & 174 & 0.84 & $7.82\times10^{-5}$\\
	22 & 203 & 0.84 & $2.93\times10^{5}$\\
	22 & 209 & 0.87 & $1.33\times10^{5}$\\
	23.15 & 232 & 0.84 & $9.54\times10^{1}$\\
	24 & 246 & 0.84 & $2.26\times10^{2}$\\
	\hline\hline
	\end{tabular}
\end{table}

\subsection{Freezing at Vapor-Liquid Interfaces\label{section:interface}}
As evident in Table~\ref{table:rates} and Fig.~\ref{fig:rates}, a crossover exists between the bulk-dominated crystallization at $\lambda\ge22$ and the surface-enhanced crystallization at $\lambda = 21$. Table~\ref{table:ratios} summarizes $R_{\text{bulk}}/R_{\text{film}}$ for the films considered in this work. The sensitivity of the nucleation rate to the presence or absence of an interface  tends to be  stronger at lower $\lambda$'s. For $\lambda=22$, for instance, nucleation in the film is almost five orders of magnitude slower than in the bulk, while at $\lambda=21$, it is five orders of magnitude faster. Contrast this to the bulk-to-film ratios at higher $\lambda$'s in which the ratio is only around two orders of magnitude. 

Earlier studies of surface crystallization in silicon, another tetrahedral liquid, have revealed that the facilitation or suppression of crystallization in the subsurface region might depend on temperature~\cite{LiNatMater2009}. In their simulations of silicon films using the Tersoff potential~\cite{TersoffPhysRevB1989}, Li~\emph{et al.} had discovered that surface-induced crystallization is only observed for ${\zeta}>0.86$, with surface crystallization being suppressed at lower temperatures. In order to confirm that our observation of the bulk-dominated crystallization at $\lambda = 22$ and $T=203$~K is robust, we carry out another set of rate calculations at $T=209$~K, which corresponds to a relative supercooling of ${\zeta}\approx0.87$. In those calculations, we still observe the suppression of nucleation at the vapor-liquid interface.

\begin{figure}
	\centering
	\includegraphics[width=.45\textwidth]{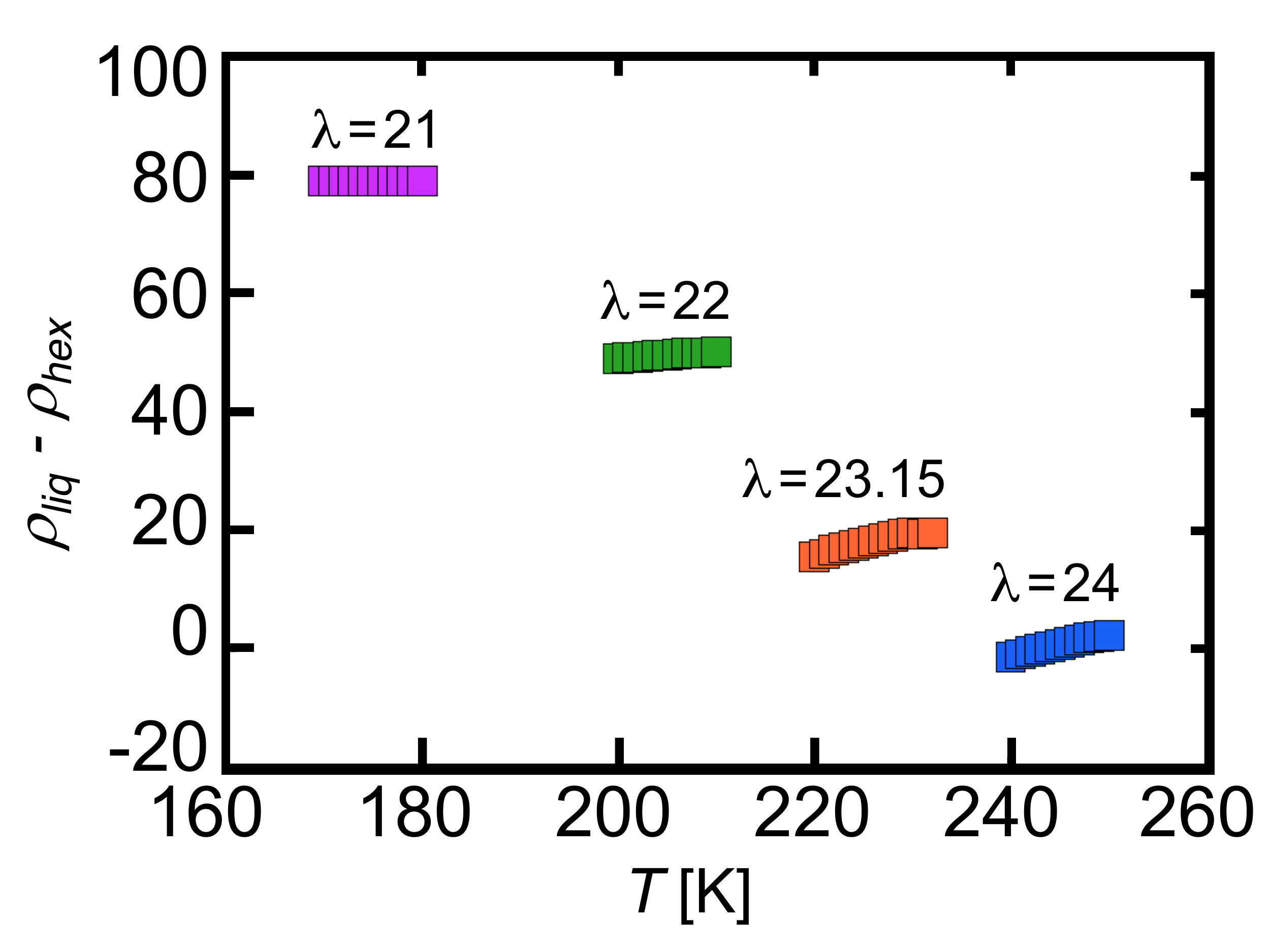}
	\caption{\label{fig:density-diff} Density difference between the hexagonal crystal and the supercooled liquid at $p=1$~bar for different $\lambda$'s. }
\end{figure}

In Section~\ref{section:introduction}, we provide a thorough discussion of different attempts for establishing a correlation between the thermodynamic properties of a material and its propensity to surface-dominated crystallization. Features such as partial wettability of a crystal by its liquid~\cite{TabazadehPNAS2002}, as well as having a negatively-sloped melting curve~\cite{LiNatMater2009} have been associated with surface-induced crystallization. Earlier studies of nucleation kinetics in freestanding thin mW films~\cite{HajiAkbariFilmMolinero2014} have demonstrated that none of the above-mentioned features can be used as predictive indicators of surface-freezing as surface crystallization is suppressed in the mW system, despite satisfying both these criteria.

The inability of  negative-slope melting curves to predict the enhancement of surface crystallization is further emphasized in this work. Fig.~\ref{fig:density-diff} depicts the density difference between the liquid and the hexagonal crystal for temperatures in the vicinity of ${\zeta}=0.845$.  For $\lambda<24$, the liquid is always denser than the crystal, with the density difference increasing upon decreasing $\lambda$. For $\lambda=24$, however, there is almost no density difference between the liquid and the crystal. The increase in liquid density upon decreasing $\lambda$ is consistent with the widening of the first peak of $g(r)$ in Fig.~\ref{fig:RDF} and is due to the merging of the first and second nearest neighbor shells at low tetrahedralities. It must be noted that no correspondence exists between the crossover in density difference at $\lambda = 24$, and the crossover of surface crystallization kinetics at $\lambda = 21$. From a molecular perspective, it is difficult to understand how a simple variable such as density difference would capture the effect of structural intricacies of the interfacial region on a phenomenon as complex as surface freezing. 
%However, it might be possible to establish correlations between the relative density difference between the liquid and the crystal, i.e.,~$(\rho_{\text{liq}}-\rho_{\text{sol}})/\rho_{\text{sol}}$, and the propensity to freeze at vapor-liquid interfaces, at least for certain classes of materials that are structurally similar, such as the tetrahedral liquids considered in this work. 

\begin{figure}
	\centering
	\includegraphics[width=.45\textwidth]{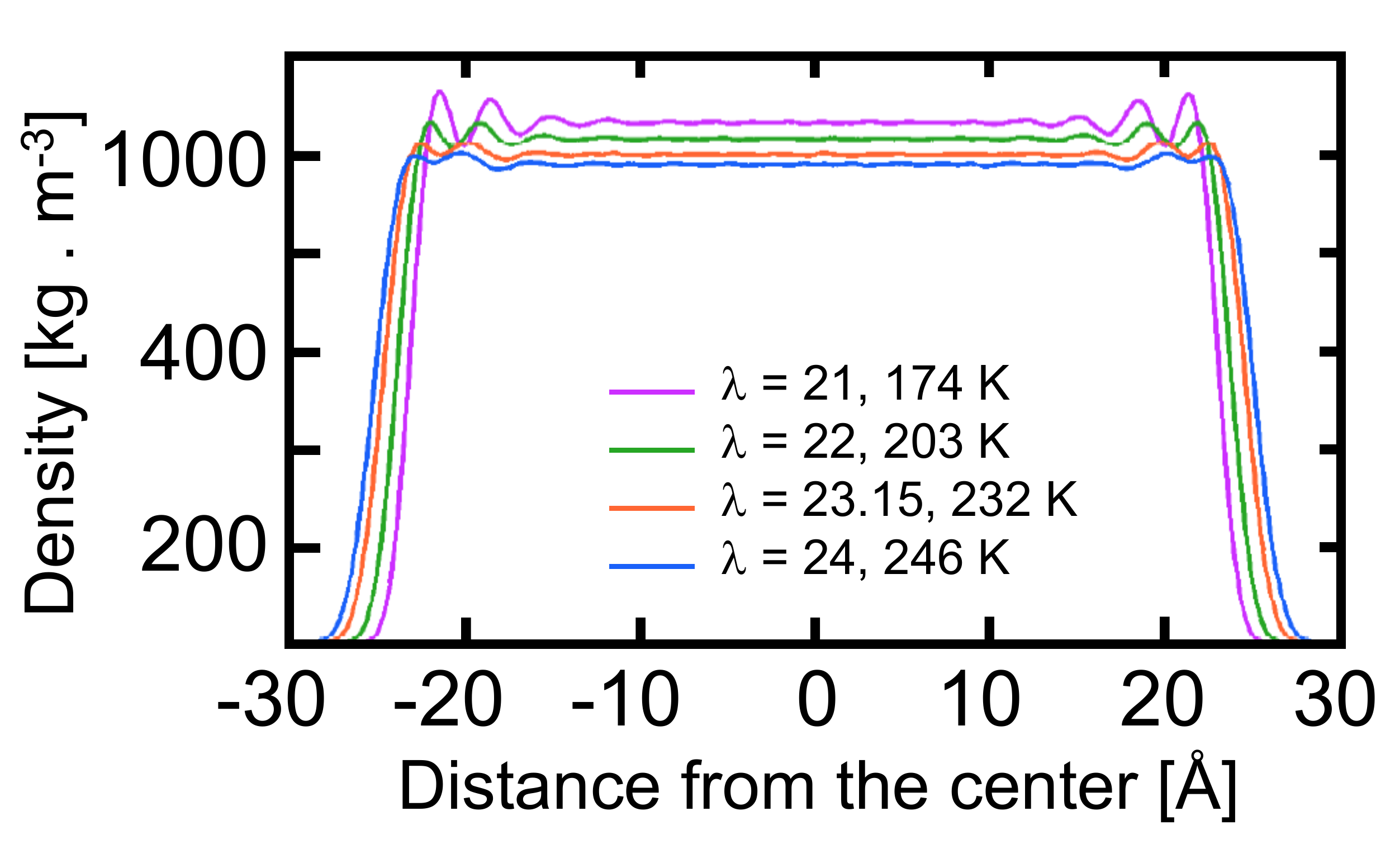}
	\caption{\label{fig:density-profiles}Density profiles across the 5-nm films at ${\zeta}=0.845$.}
\end{figure}

\begin{figure}
	\centering
	\includegraphics[width=.5\textwidth]{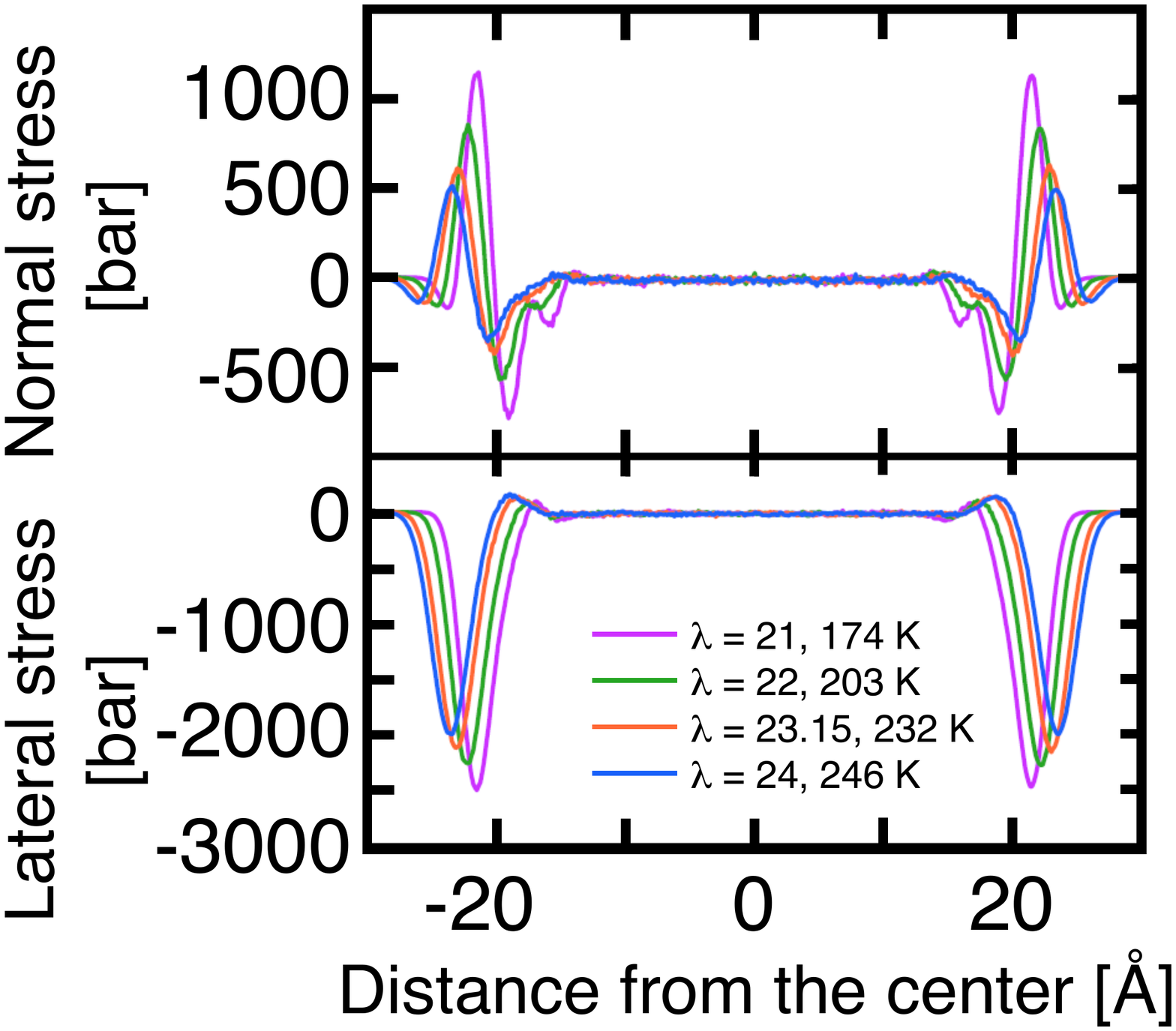}
	\caption{\label{fig:stress-profiles}(a) Normal and (b) lateral stress profiles across the 5-nm films at ${\zeta}=0.845$.}
\end{figure}

Another feature of a material that can potentially impact the crystallization kinetics close to the vapor-liquid interface is the microstructure of the interfacial region induced as a result of confinement. In order to understand the effect of $\lambda$ on the molecular structure of the interfacial region, we compute density (Fig.~\ref{fig:density-profiles}) and lateral and normal stress profiles (Fig.~\ref{fig:stress-profiles}) across the 5-nm films, using the approach discussed in Ref.~\cite{HajiAkbariJCP2014}. The films tend to become more structured as $\lambda$ decreases. This is evident in the emergence of more peaks in the density profile (Fig.~\ref{fig:density-profiles}), and more prominently, a second peak in the normal stress profile (Fig.~\ref{fig:stress-profiles}). Such enhanced structuring can, indeed, make a film more amenable to crystallization as the most structured film, i.e.,~$\lambda=21$, is the very film in which surface crystallization is enhanced. This picture is, however, incomplete as the emergence of the above-mentioned peaks first occurs at $\lambda=22$, and it is not clear how and why such an enhanced structuring at $\lambda = 22$ does not translate into the facilitation of  crystallization at the surface. Note that the thickness of the interfacial region, defined as the region with anisotropic stress tensor, is virtually insensitive to $\lambda$ and is around 12-13~\AA~in all the films considered in this work. 

Another peculiarity that is noted in Table~\ref{table:ratios} is the especially strong suppression of crystallization at $\lambda=22$. We can explain this anomaly by the following two considerations. First, films are thinner at lower $\lambda$'s due to the increase in $\rho_{\text{liq}}$ as $\lambda$ decreases. This increases the relative share of the interfacial region in the entire volume of the film as the width of the interfacial region is almost constant. As demonstrated in Ref.~\cite{HajiAkbariFilmMolinero2014}, the suppression of surface crystallization is more pronounced in thinner films. Secondly, the interfacial region is more anisotropic at $\lambda=22$, in the sense that the difference between the lateral and normal stress is larger. Therefore, whatever feature that suppresses surface crystallization for $\lambda\ge22$ is likely to be stronger at $\lambda=22$. This will lead to a decrease in growth probabilities of large crystallites, due to the asymmetric growth of the confined crystallites discussed in Ref.~\cite{HajiAkbariFilmMolinero2014}. In this context, the side of the nucleus that is exposed to the highly anisotropic interfacial region might be less likely to absorb new liquid-like molecules at $\lambda=22$  than higher $\lambda$ values. The combination of these two effects can decrease the effective volumetric nucleation rate, $R_{\text{film}}$, by larger quantities at $\lambda=22$ in comparison to higher $\lambda$'s.

\section{Conclusions\label{section:conclusions}}
In this work, we investigate the effect of $\lambda$, the tetrahedrality parameter, on the kinetics of crystal nucleation in a family of Stillinger-Weber potentials, with $\epsilon$ and $\sigma$ taken from the mW potential, but with $21\le\lambda\le24$. We observe that the nucleation rate is a strong function of $\lambda$ and changes by approximately 48 orders of magnitude upon changing $\lambda$ from 21 to 24 at a relative supercooling of ${\zeta}=0.845$. By computing self-diffusivities at different $\lambda$'s, we conclude that the kinetic pre-factor in CNT is virtually insensitive to $\lambda$, and the change in rate in predominantly a consequence of the change in the nucleation barrier. We also use thermodynamic integration to estimated $|\Delta\mu|$, the thermodynamic driving force, for different $\lambda$'s and observe that $|\Delta\mu|$, increases upon increasing $\lambda$. By assuming the validity of  classical nucleation theory, we demonstrate that the observed change in $R$ can, for the most part, be accurately explained by the corresponding change in $|\Delta\mu|$, suggesting that any possible change of $\gamma_{ls}$ with~$\lambda$ is  too small to affect the nucleation kinetics. We also examine the role of vapor-liquid interfaces on freezing, and observe a crossover between the bulk-dominated freezing at $\lambda\ge22$, and surface-dominated freezing at $\lambda$ = 21. We observe that the interfacial region becomes more structured at lower $\lambda$'s. However, this enhanced structuring starts  at $\lambda=22$, which does not coincide with the observed crossover into surface-induced crystallization at $\lambda=21$. The existence of a negatively-sloped melting curve is not predictive either, as the liquid is denser than the crystal for $\lambda<24$. This underscores the difficulty of identifying the true cause of surface-induced crystallization, as the existing heuristics-- i.e.,~the negative-slope melting curve, and  the partial wettability of the crystal-- seem to be incapable of explaining the trends observed here and in Ref.~\cite{HajiAkbariFilmMolinero2014}. 

This current work is among the first to systematically investigate the effect of certain features of a water model on its thermodynamic and kinetic properties. This approach can also provide us with a more fundamental perspective of phase transitions in aqueous systems, and how they are affected by water anomalies. A similar approach has been recently used to investigate the effect of other features of water models, such as the hydrogen bond flexibility, on the existence of a second liquid-liquid critical point in the ST2 model~\cite{SciortinoPRL2015}. Such studies will collectively enrich our knowledge of water and its structural, kinetic and thermodynamic peculiarities.

\acknowledgments
P.G.D. gratefully acknowledges the support of the National Science Foundation (Grant No.~CHE-1213343), the Carbon Mitigation Initiative at Princeton University (CMI), and the Argentine National Agency for Scientific and Technological Promotion (Grant. PICT-Ra\'{i}ces PICT 2010-1291). 
M.P.L. thanks financial support from ANPCyT - FONCyT (PICT 2010 N$^{\circ}$1291).
M.M.G. acknowledges a fellowship from CONICET. M.P.L. is a member of CONICET.
These calculations were  performed on the Terascale Infrastructure for Groundbreaking Research in Engineering and Science (TIGRESS) at Princeton University. This work also used the Extreme Science and Engineering Discovery Environment (XSEDE), which is supported by National Science Foundation grant number ACI-1053575.  

\bibliographystyle{rsc}
\bibliography{References}

\end{document}